\documentclass[prd,twocolumn,showpacs,preprintnumbers,nofootinbib]{revtex4}
\usepackage{amsfonts,amsmath,amssymb}
\usepackage{graphicx}
\usepackage{color}
\usepackage[
  separate-uncertainty = true,
  multi-part-units = repeat
]{siunitx}

\usepackage[plainpages=false, colorlinks=true, anchorcolor=blue, linkcolor=blue, citecolor=blue, bookmarks=false]{hyperref}
\usepackage{natbib}
\pdfoutput=1

\begin{document}
\title{Galaxy cluster hydrostatic masses using Tolman-Oppenheimer-Volkoff equation}
\author{Sajal \surname{Gupta}$^1$}
\altaffiliation{E-mail: sg15ms084@iiserkol.ac.in}
\author{Shantanu  \surname{Desai}$^2$} \altaffiliation{E-mail: shntn05@gmail.com}
\affiliation{$^{1}$Department of Physical Sciences, IISER-Kolkata, Mohanpur, West Bengal-741246, India}
\affiliation{$^{2}$Department of Physics, Indian Institute of Technology, Hyderabad, Telangana-502285, India}

\begin{abstract}
Motivated by previous studies in literature  about the potential importance of  relativistic corrections to galaxy cluster hydrostatic masses, we  calculate  the masses of 12 relaxed  clusters (with Chandra X-ray data) using the Tolman-Oppenheimer-Volkov  (TOV) equation of hydrostatic equilibrium and the ideal gas equation of state. Analytical formulae  for gas density and temperature profiles for these clusters, previously     derived by Vikhlinin et al~\cite{Vikhlinin06}  were used to obtain these masses. We compare the TOV-based masses with those obtained using the corresponding Newtonian equation of hydrostatic equilibrium.  We find  that the fractional relative difference between the two masses are negligible, corresponding to  $\sim \mathcal{O}(10^{-5})$.

\end{abstract}
\maketitle
\section{Introduction}
Ever since the discovery of the accelerating universe, a large number of observational surveys have been undertaken, using state of the art multi-CCD imagers to precisely pin down the observational characteristics of this acceleration~\cite{Huterer}. The physical cause of this acceleration is currently unknown and all theoretical explanations are currently subsumed under the moniker ``dark energy''. The most prosaic explanation is a fluid characterized by the equation of state $P=w\rho$, with $w<-1/3$~\cite{Turner98} and $w=-1$ corresponds to the Cosmological Constant~\cite{Weinberg}. However, other possibilities include 
addition of scalar fields, modified theories of gravity or  inhomogenous solutions to Einstein's equations for Cosmology. See Ref.~\cite{Huterer,book} (and references therein) for a review of all theoretical possibilities for the accelerating universe.

One way to distinguish between modified gravity and dark energy is using galaxy clusters. Galaxy clusters are the most massive collapsed objects and have been used to obtain pin down the dark energy equation of state using both cluster counts and gas mass fraction~\cite{Voit,Allen,Kravtsov,Vikhlininrev,Rapetti19}. Galaxy clusters are also wonderful laboratories for constraining fundamental physics, such as  bounding the neutrino mass~\cite{Takada,Carbone}, graviton mass~\cite{Desai18,Gupta,Gupta19}, or limits on primordial non-gaussianity~\cite{Shandera}. One reason for galaxy clusters playing such an important role for cosmology   in the last decade, is due to the large number of new discoveries,  courtesy dedicated optical, X-ray, and microwave surveys.

One crucial ingredient in  using galaxy clusters as  laboratories for the above studies is the  accurate determination of their masses with negligible error bars. Galaxy cluster masses have been traditionally  determined using three methods: velocity dispersions, X-ray profiles or SZ observations assuming hydrostatic equilibrium, and gravitational lensing. These techniques are extensively reviewed in Refs.~\citep{Sarazin86,Voit, Allen,Ettori,Kravtsov,Hoekstra,Vikhlininrev}. All these works  (except those involving gravitational lensing) derive the cluster mass in the Newtonian approximation. Therefore, all cosmological results from clusters using SZ, X-rays and velocity dispersions are determined under the premise of Newtonian gravity.

The first  detailed study of  general relativistic corrections   to  galaxy cluster masses in the Newtonian limit was done by Bambi~\cite{Bambi}. He pointed out that for the Kottler space-time, which is the spherically symmetric solution to Einstein's vacuum field equations with a cosmological constant~\cite{Perlick},  the effective cluster mass is given  in the Newtonian limit by $M_N -\frac{8}{3}\pi r^3 \rho_{\Lambda}$, where $M_N$ is the Newtonian  cluster mass obtained from velocity dispersions or  hydrostatic equilibrium equation (HSE, hereafter and introduced in Eq.~\ref{eq:N})  and  $\rho_{\Lambda}$ is the energy density in the Cosmological constant.
The gravitational lensing-based mass estimates do not have such a correction. He estimated that for galaxy cluster masses of $\sim 10^{13} M_{\odot}$, the discrepancy with lensing-based mass estimates could be upto 40\%.  He also proposed a way to detect the effects of non-zero $\Lambda$, using galaxy cluster masses at 1 Mpc.
 Bisnovatyi-Kogan and  Chernin studied the dynamics of the VIRGO cluster
by adding a repulsive force  due to the cosmological constant to the Newtonian  gravitational force~\cite{Chernin12}, and showed that the radial extent and average density of dark matter haloes is determined by the energy density in the Cosmological constant. In Ref.~\cite{Chernin13}, the effective gravitating mass of the Coma cluster  was obtained using the prescription in Ref.~\cite{Bambi}, and found to be $\sim 2.4 \times 10^{15} M_{\odot}$, compared to the Newtonian mass estimate of $\sim 6.2 \times 10^{15} M_{\odot}$. 

Here we try to evaluate the galaxy cluster masses using  the  general relativistic   HSE, and consider how
this changes galaxy cluster masses determined using X-ray profiles under the premise of spherical symmetry. 
For relaxed galaxy clusters, the starting point for estimating the cluster mass is to posit the Newtonian HSE for spherical symmetry:
\begin{equation}
    \frac{dP}{dr} = \frac{-G M\rho}{r^2}
    \label{eq:N}
\end{equation}
This equation is also used to determine the stellar structure. For an ideal gas equation of state, this can be turned around to determine the HSE mass $M(r)$ in terms of the gas temperature $T(r)$ and density  profile $\rho(r)$~\cite{Allen,Ettori} :
\begin{equation}
M (r) = -\frac{k_bT r}{G\mu m_p}\left(\frac{d\ln\rho (r)}{d\ln r}+ \frac{d\ln T}{d\ln r}\right),
\label{eq:1}
\end{equation}
Since dark matter is assumed to be pressure-less~\cite{Sartoris},  we can ignore its contribution in the hydrostatic equilibrium equation.
This equation is used in a large number of studies to determine the total X-ray cluster mass, mass-concentration relations, derivation of  mass proxies from observables, etc.~\cite{Reiprich02,Vikhlinin06,Capasso16}. Such inferences are in turn used  to determine cluster cosmology or other fundamental physics parameters~\cite{Vikhlinin09}. 

 In this work, we examine the effect of GR corrections to Eq.~\ref{eq:1}, where it gets superseded by  the Tolman-Oppenheimer-Volkoff (TOV) equation. Our main goal is  check if the TOV equation causes appreciable differences to Newtonian HSE masses. The data sample we use for such a pilot study is a sample of 12 clusters obtained using pointed and  archival Chandra and ROSAT observations by Vikhlinin et al~\cite{Vikhlinin05,Vikhlinin06}. We have previously used this sample to bound the graviton mass~\cite{Gupta19}.

Comparison of galaxy cluster mass
estimates from different observational proxies has shown that the  X-ray masses are systematically underestimated 
with respect to the weak lensing masses~\citep{Zhang08,Mahdavi,Vonderlinden14,Comalit,Donahue}. Although more prosaic astrophysical explanations, such as non-thermal pressure support~\cite{Ettori} have been proposed to explain these discrepancies, another aim of this work is to see if GR corrections to Newtonian HSE can help alleviate some of these problems or cause any important systematic effects.

The outline of this manuscript is as follows. We review the TOV equation in Sect.~\ref{sec:TOV} and indicate how the HSE mass can written  in terms of the gas temperature  and density profile. Our analysis and results can be found in Sect.~\ref{sec:results}.  We conclude in Sect.~\ref{sec:conclusions}.

\section{TOV equation}
\label{sec:TOV}
The TOV~\cite{Tolman,TOV} equation  can be derived  by positing a  generic metric valid for a static isotropic spacetime and assuming a perfect fluid for the stress-energy tensor. If we assume spherical symmetry, implying that the pressure $P$ and density $\rho$ are functions of only the radial coordinate $r$, then one obtains the TOV equation (cf. GR textbooks such as Refs.~\cite{Weinberg72, Glendenning} for a full  derivation):
\begin{widetext}
\begin{equation}
\frac{dP}{dr} = -\frac{G}{r^2}\left[M(r) + \frac{4\pi r^3 P(r)}{c^2}\right]\left[\rho(r)+ \frac{P(r)}{c^2}\right]\left[1-\frac{2 G M(r)}{c^2 r}\right]^{-1}
\label{eq:TOV}
\end{equation}
\end{widetext}

This equation represents the reduction of Einstein's equations for the interior of a spherical, static object.  This equation is the GR analog of Eq.~\ref{eq:N}. The TOV equation is routinely used to determine the mass-radius relation of neutron stars as well as the maximum mass, once  an  equation of state $P(\rho)$ is specified~\cite{Lattimer}. The main difference between the TOV equation and its Newtonian counterpart (Eq.~\ref{eq:N}) is that in the latter, pressure  supports the object against gravity, whereas in the TOV equation, pressure gravitates.  That is why in very strong gravitational fields, pressure expedites gravitational collapse,  ultimately causing the object to collapse to  a black hole~\cite{Thorne65}.

In this work, we use the TOV equation to calculate the HSE mass of a sample of galaxy clusters (observed  using Chandra X-ray data) and compare these mass estimates with the corresponding Newtonian value. We should point out that the assumption of spherical symmetry is not always valid for clusters and they are known to be prolate~\cite{Jing}. However,  spherical symmetry is commonly assumed to obtain galaxy cluster mass  from Newtonian HSE (eg. Ref.~\cite{Vikhlinin06,Ettori}), and therefore we evaluate  its relativistic incarnation under the same premise of spherical symmetry to estimate the change in X-ray mass.

\section{Analysis and Results}
\label{sec:results}
Vikhlinin et al~\citep{Vikhlinin06} (V06, hereafter) derived the density and temperature profiles for a  total of 13 nearby relaxed galaxy clusters, from pointed as well as archival Chandra and ROSAT  observations. These profiles are valid upto approximately  1 Mpc. We have previously used this data for  12 of these clusters
to obtain a bound on the graviton mass~\citep{Gupta19}, and the same data has been used to constrain a plethora of modified gravity theories and alternatives to $\Lambda$CDM model~\cite{Rahvar,Khoury17,Bernal,Ng,Edmonds,Hodson17}.
We use the same data for these 12 clusters for this analysis. Note that similar to Ref.~\cite{Gupta19}, we omitted  USGC 2152 (among the 13 clusters) from our analysis, as the pertinent data was not available to us.

To obtain the GR-based HSE mass, we first  invert Eq.~\ref{eq:TOV} to derive the mass at a given radius $M(r)$  in terms of the temperature ($T(r)$) and pressure profiles ($P(r)$):
\begin{widetext}
\begin{equation}
M(r) = \left[\frac{dP}{dr}+ \frac{4\pi G r\rho(r)P(r)}{c^2}\left(1+\frac{P(r)}{\rho(r) c^2}\right)\right]\left[\frac{2G}{c^2r}\frac{dP}{dr} - \frac{G\rho(r)}{r^2}\left(1+\frac{P(r)}{\rho(r) c^2}\right)\right]^{-1}
\label{eq:TOVmass}
\end{equation}
\end{widetext}

Similar to V06, we then assume an ideal gas equation of state $P=\rho K_b T/\mu m_p G$, where $m_p$ is the mass of the proton,  $\mu$ is the mean molecular weight of the cluster in a.m.u. $\sim 0.6$~\cite{Vikhlinin05}. We then use the ideal gas law to plug in $dP/dr$ in terms of $\frac{dT}{dr}$ and $\frac{d\rho}{dr}$ in Eq.~\ref{eq:TOVmass}. Analytic formula for $T(r)$ and $\rho(r)$ for these clusters have been provided in Eq.~3 and Eq.~6 of  V06 (and also reproduced  in Refs.~\cite{Rahvar,Gupta19}).  For brevity, we do not regurgitate the same formulae here, and details of these models can be found in the above references. The analytic formulae for $\frac{dT}{dr}$ and $\frac{d\rho}{dr}$ are also available in Appendix C of Ref.~\cite{Rahvar}, which we have used to solve Eq.~\ref{eq:TOV}.

\begin{table}[h]
\begin{tabular}{|l|c|c|c|} \hline
Cluster & $r_{500}$ & $M_{Newt}$ & $[\Delta M(r)/M(r)]_{r=r_{500}}$ \\
&  (kpc) & $(M_{\bigodot})$ & (\%) \\
\hline
A133 & 1007 $\pm$ 41 & $4.1 \times 10^{14}$ & $4.3 \times 10^{-3}$ \\ \hline
A262 & 650 $\pm$ 21 &  $8.3 \times 10^{13}$ & $1.5 \times 10^{-3}$ \\ \hline
A383 & 944 $\pm$ 32 & $3.2 \times 10^{14}$ & $3.8 \times 10^{-3}$ \\ \hline
A478 & 1337 $\pm$ 58 & $7.4 \times 10^{14}$ & $6.4 \times 10^{-3}$ \\ \hline
A907 & 1096 $\pm$ 30 & $4.4 \times 10^{14}$ & $4.7 \times 10^{-3}$ \\ \hline
A1413 & 1299 $\pm$ 43 &  $9.9 \times 10^{14}$ & $8.3 \times 10^{-3}$ \\ \hline
A1795 & 1235 $\pm$ 36 & $5.8 \times 10^{14}$ & $5.2 \times 10^{-3}$ \\ \hline
A1991 & 732 $\pm$ 33 & $1.4 \times 10^{14}$ & $2.1 \times 10^{-3}$ \\ \hline
A2029 & 1362 $\pm$ 43 & $8.6 \times 10^{14}$ & $7.1 \times 10^{-3}$ \\ \hline
A2390 & 1416 $\pm$ 48 & $1.3 \times 10^{15}$ & $1.1 \times 10^{-2}$ \\ \hline
MKW 4 & 634 $\pm$ 28 & $7.5 \times 10^{13}$ & $1.9 \times 10^{-3}$ \\ \hline
RXJ1159 & 700 $\pm$ 57 & $1.21 \times 10^{14}$ & $1.3 \times 10^{-3}$ \\ \hline
\end{tabular}
\caption{Newtonian HSE mass (third column) as well as the fractional difference  between the TOV and Newtonian Mass (fourth column) evaluated at $r_{500}$ for 12 clusters in in V06. Data for $r_{500}$ was obtained from V06. As we can see, the fractional differences in the two mass estimates are negligible.}
\label{tab:results}
\end{table}

\begin{figure*}
 \begin{minipage}{\textwidth}
    \centering
    \includegraphics[width=1.0\textwidth]{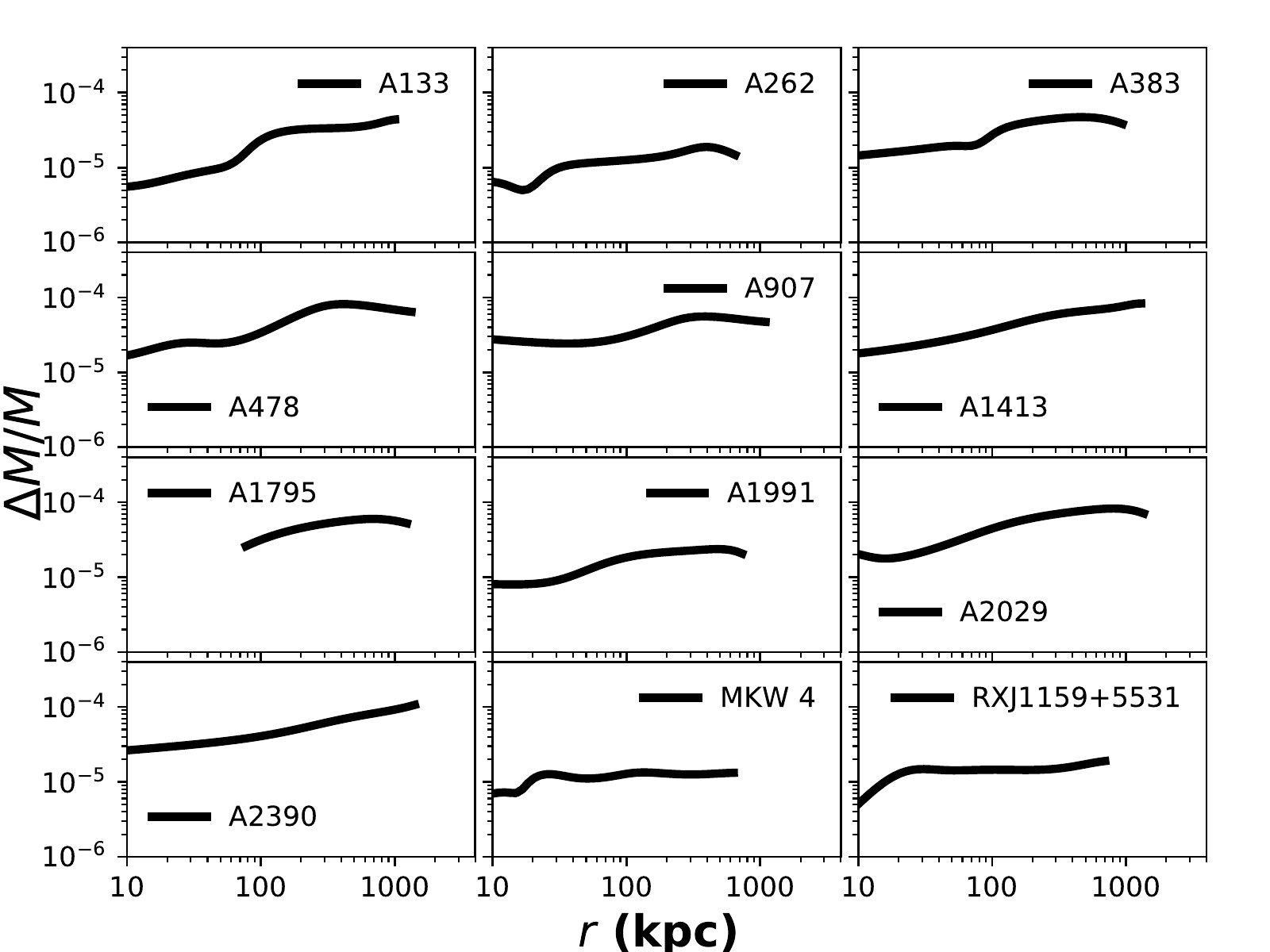}
    \caption[Caption for LOF]{Fractional mass difference between the TOV  and Newtonian HSE mass (evaluated using Eqns.~\ref{eq:TOV} and \ref{eq:N} respectively) for    12 relaxed galaxy clusters as a function of distance from the cluster center ($r$).\footnotemark{} The maximum difference between them is about 0.01\% at around 1~Mpc. Note that the difference is much smaller than the uncertainties (not shown here) in the mass estimates, which are about 10\%~\cite{Vikhlinin06}.}
    \footnotetext {For A1795, we have plotted these differences starting from $\sim$ 80 kpc, instead of 10 kpc. Below this value, we were getting negative values for both the masses. This could be  because the coefficients in the formulae for the  temperature and density profiles for this cluster are not valid below 100 kpc. In V06 also, all the profiles for this cluster have been shown starting from 100 kpc.}
    \label{fig:Rel.error}
    \end{minipage}
\end{figure*}


We now use all these  terms to solve for the GR HSE mass in Eq.~\ref{eq:TOVmass}, and compare it to the Newtonian mass (obtained using Eq.~\ref{eq:N}). We plot the fractional relative  difference in both the masses as a function  of distance from the cluster center ($r$). These deviations are shown in Fig.~\ref{fig:Rel.error} and the corresponding values at $r_{500}$ for all the 12 clusters are tabulated in Table~\ref{tab:results}. We can see that the median  difference is $\sim \mathcal{O}(10^{-5})$. The maximum difference reaches about 0.01\% for the A2390 cluster.  These are much smaller than the estimated uncertainties in Newtonian HSE masses in V06, which are about 10\%. 

Therefore, differences between the TOV and Newtonian masses are completely negligible and cannot account for some of the known discrepancies between X-ray and lensing-based mass estimates discussed in Ref.~\cite{Ettori}.

\section{Conclusions}
\label{sec:conclusions}
A large number  of galaxy clusters have been observed in X-rays since the 1970s~\cite{Sarazin86}. For relaxed galaxy clusters, there is a vast amount of literature which has used the Newtonian hydrostatic equilibrium equation to obtain the masses~\cite{Allen}. These masses have been used to obtain cosmology results, derive scaling relations based on various observables or compare to SZ and lensing-based mass estimates.

In the past decade, some authors have pointed that general relativistic corrections based on the Kottler  metric to galaxy cluster mass estimates obtained under the premise of Newtonian gravity,  can cause significant differences to the hydrostatic mass estimates for galaxy clusters  depending on the distance from the center~\cite{Bambi,Chernin12,Chernin13}. Here, we carry out a pilot study  to investigate the impact of using the TOV equation to estimate the hydrostatic mass of relaxed galaxy clusters. The TOV equation (cf. Eq.~\ref{eq:TOV}) is the GR analog of the usual (Newtonian) hydrostatic equilibrium equation  (cf. Eq.~\ref{eq:N}) for spherically symmetric systems,  and is usually used to describe the structure and mass-radius relationship of neutron stars~\cite{Lattimer}. This is the first study (to the best of our knowledge) which investigates the impact of  TOV equation for galaxy clusters.

The data sample we used for this study consists of about 12 clusters observed with archival and pointed Chandra and ROSAT observations~\cite{Vikhlinin05,Vikhlinin06}. For each of these 12 clusters, analytic formulae for gas density and temperature profiles  have been made available~\cite{Vikhlinin06}. This dataset has been used in the literature  to test a plethora of modified gravity theories, and most recently also used by us  to constrain the graviton mass~\cite{Gupta19}. For all the 12 clusters, we derive the TOV based mass and compare it to the Newtonian mass as a function of distance from the galaxy cluster. These are shown in Fig.~\ref{fig:Rel.error} and also tabulated at $r_{500}$ in Table~\ref{tab:results}. We  see that the differences in mass estimates are negligible ($\sim \mathcal{O} (10^{-5})$) and the maximum difference we find is for A2390 of about 0.01\% and much smaller than the errors in mass estimates.

Therefore for all practical purposes, one can safely use the Newtonian equation of hydrostatic equilibrium for any dynamical analysis with galaxy clusters.

\begin{acknowledgements}
Sajal Gupta is  supported by a DST-INSPIRE fellowship. We are grateful to Cosimo Bambi for comments on the manuscript and to Alexey Vikhlinin for providing us the data from V06.
\end{acknowledgements}
\bibliography{main}
\end{document}